\documentclass[lettersize,journal]{IEEEtran}
\usepackage{amsmath,amsfonts}
\usepackage{algorithm}
\usepackage{array}
\usepackage[caption=false,font=normalsize,labelfont=sf,textfont=sf]{subfig}
\usepackage{textcomp}
\usepackage{stfloats}
\usepackage{url}
\usepackage{verbatim}
\usepackage{graphicx}
\usepackage{algpseudocode}
\usepackage{subfig}
\usepackage{acronym}
\usepackage{cite}
\usepackage[none]{hyphenat} 
\acrodef{rf}[RF]{radio frequency}
\acrodef{irs}[IRS]{intelligent reflecting surfaces}
\acrodef{pls}[PLS]{physical layer security}
\acrodef{ga}[GA]{genetic algorithm}
\acrodef{dco-ofdm}[DCO-OFDM]{direct current optical frequency division multiplexing }
\acrodef{led}[LED]{light-emitting diode}
\acrodef{los}[LoS]{line-of-sight}
\acrodef{nlos}[NLoS]{non-line-of-sight}
\acrodef{pd}[PD]{photo-detector}
\acrodef{cir}[CIR]{channel impulse response}
\acrodef{cfr}[CFR]{channel frequency response}
\acrodef{isi}[ISI]{intersymbol interference}
 \acrodef{snr}[SNR]{signal-to-noise ratio}
 \acrodef{es}[ES]{exhaustive search}

\def\BibTeX{{\rm B\kern-.05em{\sc i\kern-.025em b}\kern-.08em
    T\kern-.1667em\lower.7ex\hbox{E}\kern-.125emX}}
\usepackage{balance}
\begin{document}

\title{Leveraging IRS Induced Time Delay for Enhanced Physical Layer Security in VLC Systems }
\author{ \vspace{-1ex}Rashid Iqbal,~\IEEEmembership{Graduate Student Member,~IEEE}, Mauro Biagi,~\IEEEmembership{Senior Member,~IEEE}, Ahmed Zoha,~\IEEEmembership{Senior Member,~IEEE}, Muhammad Ali Imran,~\IEEEmembership{Fellow,~IEEE}, and Hanaa Abumarshoud,~\IEEEmembership{Senior Member,~IEEE}
\thanks{R. Iqbal, A. Zoha, M. A. Imran, and H. Abumarshoud are with the James Watt School of Engineering, University of Glasgow, Glasgow, G12 8QQ, UK. (e-mail: r.iqbal.1@research.gla.ac.uk, \{Ahmed.Zoha, Muhammad.Imran, Hanaa.Abumarshoud\}@glasgow.ac.uk).}
\thanks{M. Biagi is with the Department of Information, Electrical, and Telecommunication (DIET) engineering, “Sapienza” University of Rome, Via Eudossiana 18, 00184 Rome, Italy (e-mail: mauro.biagi@uniroma1.it).}}

\maketitle

\begin{abstract}
Indoor visible light communication (VLC) is considered secure against attackers outside the confined area where the light propagates, but it is still susceptible to interception from eavesdroppers inside the coverage area. A new technology, intelligent reflecting surfaces (IRS), has been recently introduced, offering a way to control and steer the light propagation in VLC systems. This paper proposes an innovative approach for enhancing the physical layer security (PLS) in IRS-assisted VLC by utilising the effect of the time delay induced by the IRS reflections. More specifically, we consider how to allocate the IRS elements in a way that produces destructive intersymbol interference at the eavesdropper's location while enhancing the signal reception at the legitimate user. Our results show that, at a fixed light-emitting diode power of 3W, the secrecy capacity can be enhanced by up to 253\% at random positions for the legitimate user when the eavesdropper is located within a 1-meter radius of the LED. We show that careful allocation of the IRS elements can lead to enhanced PLS even when the eavesdropper has a more favourable position than the legitimate user.
\end{abstract}

\section{Introduction}
\IEEEPARstart{V}{isible} light communication (VLC) is gaining recognition as an innovative approach to support and complement established \ac{rf} communication networks \cite{Ref2}. Operating in the spectrum between 400 THz and 790 THz, VLC offers many advantages, including multi-Gbps wireless transmission, high data density, interference mitigation, beam steering, easy integration with lighting functionalities, and cost-effectiveness \cite{Ref3}. Despite these benefits, VLC implementation still faces some challenges related to the limited communication range, signal dropouts in obstructed environments, and security issues \cite{Ref3}.  VLC systems have the ability to utilise advanced beamforming and jamming techniques which are also used in \ac{rf} communications for improving the \ac{pls}. Moreover, the unique characteristics of light, such as its inability to penetrate opaque objects, provide a clear security advantage compared to \ac{rf} communications \cite{Ref55}.

\Ac{irs} technology, which utilises passive elements that can reflect and redirect light signals, offer a new way to enhance system performance in VLC systems. In the literature, the primary focus of \ac{irs}-assisted VLC  has been on maximising sum rate \cite{ Ref7}, mitigating path blockages \cite{Ref8}, and enhancing \ac{pls} \cite{Saifaldeen,Ref9, icc2023}. The work in \cite{Saifaldeen} integrates \ac{irs} with beamforming to dynamically optimise secrecy capacity based on deep reinforcement learning using the deep deterministic policy gradient algorithm. An \ac{irs} elements' allocation method is proposed in \cite{Ref9} to boost the secrecy capacity in non-orthogonal multiple access VLC systems, while \cite{icc2023} proposes a \ac{pls} approach based on employing a random \ac{irs} activation vector that is shared in advance with the legitimate users to hinder the eavesdroppers' ability to equalise the channel correctly.  

The mentioned VLC studies that feature \ac{irs} for the enhancement of \ac{pls} neglect an important practical consideration related to \ac{irs} deployment, namely, the time delay introduced by the \ac{irs} reflected paths, with an assumption of simultaneous arrival for all multipath signals \cite{Saifaldeen,Ref9, icc2023}. The work in \cite{Ref14} is not applicable to \ac{pls} but explores the impact of \ac{irs}-induced time delay on the characteristics of the VLC channel in the frequency domain. The results in \cite{Ref14} demonstrate that the \ac{irs} reflections can result in either a constructive or destructive effect based on the time of arrival of the reflected signals with respect to the \ac{los}. 

This letter proposes, for the first time in the open literature, an innovative approach that exploits   \ac{irs}-induced time delay for enhancing the \ac{pls} of \ac{irs}-assisted VLC systems.  We formulate an optimisation problem aiming to maximise the achievable secrecy capacity by allocating the \ac{irs} elements in a way that produces constructive and destructive \ac{isi} patterns at the legitimate user and the eavesdropper locations, respectively. We implement an IRS elements' allocation strategy based on the \ac{ga} to show case the performance enhancement offered by the proposed approach.

\section{System Model}
\label{section2}
\begin{figure}[!t]
\centering
\includegraphics[width=\linewidth]{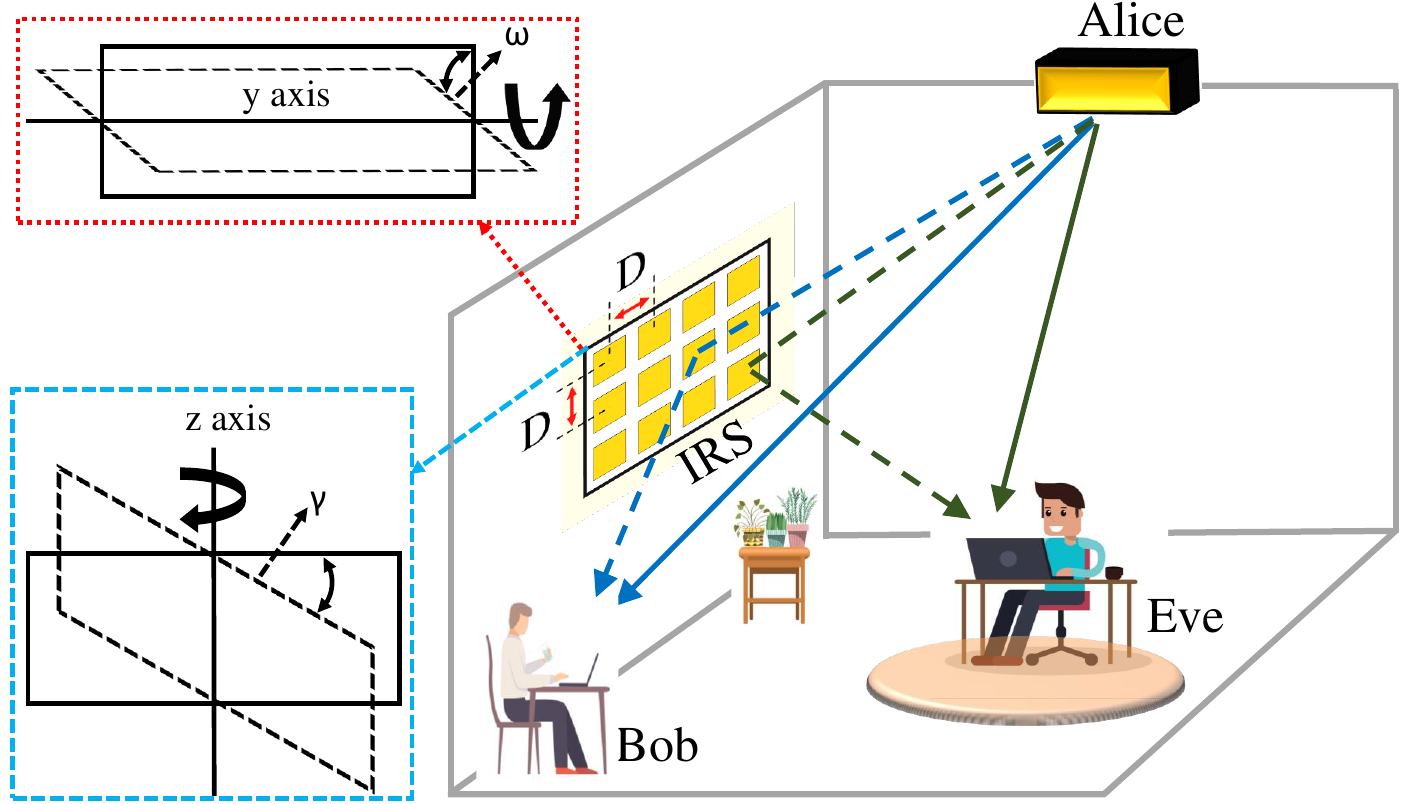}
\caption{System model of an indoor \ac{irs}-assisted VLC system with one legitimate user and one eavesdropper.}
\label{fig_a}
\end{figure}

This work considers the downlink transmission in an \ac{irs}-assisted VLC system based on \ac{dco-ofdm}. As shown in Fig. \ref{fig_a}, the system consists of a single \ac{led}, referred to as Alice, located in the center of the room and two VLC users, one is a legitimate network user, denoted as Bob, and the other is an eavesdropper, referred to as Eve, both situated at random positions within the room. Additionally, there are \(N_{\text{irs}}\) reflecting elements fixed on one room wall with a separation distance $D$ between the elements. In \ac{irs}-assisted VLC, the transmitted signal reaches the receiver via two paths: \ac{los} and \ac{nlos}. These paths are discussed in detail in the following subsections.
\setlength{\abovecaptionskip}{5pt}
\setlength{\abovecaptionskip}{22pt}

\subsection{LoS Channel}
The \ac{los} signal received from the \ac{led} to the \( k \)-th user's \ac{pd}, which we denote as \(g_{\scriptscriptstyle{k,\text{LED}}}^{\scriptscriptstyle \text{LOS}}\), follows the Lambertian model and can be represented as \cite{Ref9}:

\begin{equation}
\begin{split}
g_{\scriptscriptstyle{k,\text{LED}}}^{\scriptscriptstyle \text{LOS}} = 
\begin{cases} 
\frac{(m+1) A}{2 \pi d_{\scriptscriptstyle {k,\text{LED}}}^2} \cos^m(\phi_{\scriptscriptstyle{k,\text{LED}}}) \cos(\psi_{\scriptscriptstyle{k,\text{LED}}})  \mathbb{G}_{\scriptscriptstyle{f}} \mathbb{G}_{\scriptscriptstyle{c}}, \\
\hspace{8em} \text{for } 0 \leq \psi_{\scriptscriptstyle{k,\text{LED}}} \leq \Psi_{\scriptscriptstyle\text{FoV}} \\
0, \hspace{10.5em} \text{otherwise,}
\end{cases}
\end{split}
\end{equation}
where $k \in \{B, E\}$ is the user index, taking the value $B$ for Bob and $E$ for Eve. \( A \) represents the physical area of the \ac{pd}, \( d_{\scriptscriptstyle{k,\text{LED}}} \) denotes the Euclidean distance between the \ac{led} and \ac{pd} for the \( k \)-th user, and \( \phi_{\scriptscriptstyle{k,\text{LED}}} \) and \( \psi_{\scriptscriptstyle{k,\text{LED}}} \) are the angles of irradiance and incidence between the \ac{led} and the \ac{pd}. Moreover, \(\Psi_{\scriptscriptstyle\text{FoV}}\) denotes the field-of-veiw (FoV) of the receiever, \(\mathbb{G}_{\scriptscriptstyle f}\) represents the optical filter gain, and \( \mathbb{G}_{\scriptscriptstyle c} \) represents the optical concentrator gain which is given by:

\begin{equation}
\mathbb{G}_{c}= 
\begin{cases} 
\frac{\xi^2}{\sin^2 \Psi_{\scriptscriptstyle \text{FoV}}}, & 0 \leq \psi_{\scriptscriptstyle{k,\text{LED}}} \leq \Psi_{\scriptscriptstyle\text{FoV}}, \\
0, & \text{otherwise}
\end{cases}
\end{equation}
where \(\xi\) represents the refractive index, and \( m \) denotes the Lambertian order, which is given by:
\begin{equation}
m = -\frac{1}{\log_2(\cos(\Phi_{\scriptscriptstyle 1/2}))},
\end{equation}
 where $\Phi_{\scriptscriptstyle1/2}$ is semi-angle at half power of the \ac{led}. 

\subsection{NLoS Channel}
The NLoS channel is modeled by considering a grid of passive elements, with dimensions \(N_{x} \times N_{y}\)
, which exhibits near-perfect reflectivity, as in \cite{Ref14}. Here, \(N_{x}\) and \(N_{y}\) represent the number of elements per row and column, respectively. In \ac{rf} systems, the \ac{irs} elements can modify the phase of reflected signals to enhance signal propagation.  In VLC systems, intensity modulation is utilised, hence the phase of the signal is not relevant. Instead, the orientation of each IRS element is critical to direct the light signals toward the users accurately. This is achieved by manipulating the two rotational degrees of freedom, yaw ($\omega$) and roll angles ($\gamma$), for horizontal and vertical adjustments of the IRS element is shown in Fig. \ref{fig_a}. These adjustments ensure optimal signal direction towards users, improving signal propagation in \ac{nlos} conditions \cite{Ref15}. The specular reflection gain from the \(k\)-th user to the \(n\)-th \ac{irs} element is modeled as an additive gain, detailed in \cite{Ref15}:

\begin{equation}
\begin{split}
g_{\scriptscriptstyle{k,n,\text{LED}}}^{\scriptscriptstyle \text{NLOS}} =
\begin{cases}
\frac{\rho (m+1) A }{2 \pi (d_{\scriptscriptstyle{n,\text{LED}}} + d_{\scriptscriptstyle{k,n}})^2} \cos^m(\phi_{\scriptscriptstyle{n,\text{LED}}}) \cos(\psi_{\scriptscriptstyle{k,n}}) \mathbb{G}_{\scriptscriptstyle f}\mathbb{G}_{\scriptscriptstyle c}, \\
\hspace{8em} \text{for } 0 \leq \psi_{\scriptscriptstyle{k,n}} \leq \Psi_{\scriptscriptstyle \text{FoV}} \\
0, \hspace{10.5em} \text{otherwise,}
\end{cases}
\end{split}
\end{equation}
where \( \rho \) represents the element reflectivity. The terms \( d_{\scriptscriptstyle{n,\text{LED}}} \) and \( d_{\scriptscriptstyle{k,n}} \) represent the Euclidean distances from the \ac{led} to the \( n \)-th reflecting element and from the \( n \)-th reflecting element to the \( k \)-th user, respectively. Finally, \( \phi_{\scriptscriptstyle{n,\text{LED}}} \) and \( \psi_{\scriptscriptstyle{k,n}} \) refer to the angles of irradiance and incidence relative to the \ac{irs} element path.

This work considers realistic assumptions and accounts for the time delay in light propagation for both \ac{los} and \ac{nlos} scenarios. To accurately represent channel characteristics, we calculated the overall \ac{cir} using the following equation \cite{Ref14}:
\begin{equation}
q(t) = g_{\scriptscriptstyle{k,\text{LED}}}^{\scriptscriptstyle \text{LOS}} \delta(t-\tau_{\scriptscriptstyle{k, \text{LED}}}) + \sum_{n=1}^{N_{\scriptscriptstyle{\text{irs}}}} g_{\scriptscriptstyle{k,n,\text{LED}}}^{\scriptscriptstyle \text{NLOS}} \delta(t-\tau_{\scriptscriptstyle{k,n,\text{LED}}}),
\label{equation:cfr}
\end{equation}
where \( \delta(u) \) denotes the unit impulse function. The terms \( \tau_{\scriptscriptstyle{k,\text{LED}}} = \frac{d_{\scriptscriptstyle{k,\text{LED}}}}{c} \) and \( \tau_{\scriptscriptstyle{k,n,\text{LED}}} = \frac{d_{n,\text{LED}} + d_{k,n}}{c} \) represent signal propagation delays, and \( c \) is the speed of light.\\
To determine how time delays affect system performance across frequency ranges, we analyse this through the \ac{cfr}, which is the Fourier transform of the \ac{cir}, and is given by:
\begin{equation}
Q(f) = g_{\scriptscriptstyle{k,\text{LED}}}^{\scriptscriptstyle \text{LOS}}e^{-j2\pi f\tau_{\scriptscriptstyle{k,\text{LED}}}} + \sum_{n=1}^{N_{\scriptscriptstyle \text{irs}}} g_{\scriptscriptstyle{k,n,\text{LED}}}^{\scriptscriptstyle \text{NLOS}}e^{-j2\pi f\tau_{\scriptscriptstyle{k,n,\text{LED}}}}.
\label{equa:5}
\end{equation}
\Ac{snr} compares the power of the modulated signal to the power of the background noise at the \ac{pd} and is a direct indicator of the system's capacity to convey information accurately, which can be written as \cite{Ref16}:
\begin{equation}
\gamma(f) = \frac{2T_{\scriptscriptstyle s}E(f)P_{\scriptscriptstyle{\text{opt}}}^2 |Q(f)|^2 R_{\scriptscriptstyle{\text{pd}}}^2}{\Gamma \mathsf{k}^2 \mathcal{N} },
\end{equation}
 where \( T_{\scriptscriptstyle s} \) is the symbol period, \( E(f) \) represents a uniform power distribution, \( P_{\scriptscriptstyle{\text{opt}}} \) is the optical power transmitted by the \ac{led}, \( |Q(f)|^2 \) is the channel power gain at frequency \( f \), \(R_{\scriptscriptstyle{\text{pd}}}\) is the \ac{pd} responsivity, and \(\mathcal{N}\) is noise power spectral density at the \ac{pd}. Moreover, $\Gamma$ denotes the gap value, which quantifies the disparity between theoretical predictions and actual system performance influenced by factors such as modulation and coding schemes. The modulation scaling factor, \(\mathsf{k}\), is important for adjusting the signal amplitude within the operational constraints of the \ac{led}, optimising power usage, and minimising distortion due to signal clipping.
 
 To investigate the effect of the time delay caused by a single \ac{irs} element across different frequency ranges, which determine the received \ac{snr}, the SNR value depends on $|Q(f)|^2$ can be written as:
 \begin{align}
|Q(f)|^2 &= Q(f)Q^*(f) \notag\\
         &= (g_{\scriptscriptstyle{k,\text{LED}}}^{\scriptscriptstyle \text{LOS}})^2 + (g_{\scriptscriptstyle{k,n,\text{LED}}}^{\scriptscriptstyle \text{NLOS}})^2 + 2g_{\scriptscriptstyle{k,\text{LED}}}^{\scriptscriptstyle \text{LOS}}g_{\scriptscriptstyle{k,n,\text{LED}}}^{\scriptscriptstyle \text{NLOS}}\cos(2\pi f \Delta{\tau}),
 \end{align}
where \(*\) denotes the complex conjugate. The first and second terms represent the channel gains for the \ac{los} and \ac{nlos} paths, respectively. In contrast, the third term demonstrates how constructive and destructive interference can occur due to different time delays (\(\Delta \tau =  
\tau_{\scriptscriptstyle{k,n,{\text{LED}}}} - \tau_{\scriptscriptstyle{k,\text{LED}}}\)) between these paths. If \(\cos(2\pi f \Delta \tau) > 0\), the IRS element causes constructive interference. Likewise, if \(\cos(2\pi f \Delta \tau) < 0\), the \ac{irs} element causes destructive interference.
\section{Problem Formulation and Proposed Solution }
This study investigates the maximisation of the secrecy capacity in an \ac{irs}-assisted VLC system. In our system, we utilise the \ac{snr}  to calculate the rates of Bob and Eve. The \ac{irs} elements are strategically allocated to enhance Bob's signal through constructive \ac{isi} while simultaneously reducing Eve's signal with destructive \ac{isi}. This configuration of the \ac{irs} elements is essential in maximising the secrecy capacity of the system. The achievable rate of a \ac{dco-ofdm} with frequency selective channel is given by \cite{Ref14,Ref16} :
\begin{equation}
\begin{aligned}
R_{\scriptscriptstyle k} & =\int_0^{\frac{1}{2 T_{\scriptscriptstyle s}}} \log _2\left(1+\frac{2 T_{\mathrm{s}} E(f) P_{\scriptscriptstyle{\text{opt}}}^2 R_{\scriptscriptstyle{\text{{pd}}}}^2 |Q(f)|^2 } {\Gamma \mathsf{k}^2 \mathcal{N}}\right) d f.
\end{aligned}
\label{rate}
\end{equation}
We assume that the \ac{irs} elements are based on binary vector \( {\mathbf s}  = [s_{\scriptscriptstyle 1}, s_{\scriptscriptstyle2}, \ldots, s_{\scriptscriptstyle N_\text{irs}}]^T \), where,  \(s_{\scriptscriptstyle n}\) can only be a 0 or 1. If \(s_{\scriptscriptstyle n} = 1\), the \(n\)-th \ac{irs} element is allocated to Bob, while if \(s_{\scriptscriptstyle n} = 0\), the \(n\)-th \ac{irs} element is allocated to Eve. By inserting (\ref{equa:5}) in (\ref{rate}), the rate of Bob is given as :

{\small\begin{equation}
\begin{aligned}
R_{\scriptscriptstyle B} & =\int_0^{\frac{1}{2 T_{\scriptscriptstyle s}}} \log _2\left(1+\frac{2 T_{\scriptscriptstyle s} E(f) P_{\scriptscriptstyle{\text{opt}}}^2 R_{\scriptscriptstyle{\text{pd}}}^2}{\Gamma \mathsf{k}^2 \mathcal{N}}\right. \\
& \left.\times\left|g_{\scriptscriptstyle{B, \text{LED}}}^{\scriptscriptstyle \text{LOS}} e^{-2 \pi j f \tau_{\scriptscriptstyle{B, \text{LED}}}}+\sum_{n=1}^{N_{{\scriptscriptstyle \text{irs}}}} g_{\scriptscriptstyle{ B,n,\text{LED}}}^{\scriptscriptstyle \text{NLOS}} s_{\scriptscriptstyle n}e^{-2\pi j f \tau_{\scriptscriptstyle{ B,n,\text{LED}}}}
\right|^2\right) {d} f,
\end{aligned}
\label{rate:Bob}
\end{equation}}
and the rate of Eve is given as:
{\small
\begin{equation}
\begin{aligned}
R_{\scriptscriptstyle E} & =\int_0^{\frac{1}{2 T_{\scriptscriptstyle s}}} \log _2\left(1+\frac{2 T_{\scriptscriptstyle s} E(f) P_{\scriptscriptstyle{\text{opt}}}^2 R_{\scriptscriptstyle{\text{pd}}}^2}{\Gamma \mathsf{k}^2 \mathcal{N}}\right. \\
& \left.\times\left|g_{\scriptscriptstyle E,\text{LED}}^{\scriptscriptstyle \text{LOS}} e^{-2 \pi j f \tau_{\scriptscriptstyle{ E,\text{LED}}}}+\sum_{n=1}^{N_{\scriptscriptstyle \text{irs}}} g_{\scriptscriptstyle{ E,n,\text{LED}}}^{ \scriptscriptstyle \text{NLOS}} (1-s_{\scriptscriptstyle n}) e^{-2 \pi j f \tau_{\scriptscriptstyle{E,n,\text{LED}}}}\right|^2\right) df.
\label{Eaves:rate}
\end{aligned}
\end{equation}}\\
Since we are dealing with $N_{\scriptscriptstyle \text{irs}}$ elements, it is not straightforward to quantify the effect of delay analytically but it can be observed numerically.

As discussed in Section \ref{section2}, we assume specular reflection since a single \ac{irs} element can serve one user at a time. The secrecy capacity is the difference in data rates between Bob and Eve. The optimisation problem can be formulated as follows:
\begin{subequations}
\renewcommand{\theequation}{\theparentequation\alph{equation}}
\begin{equation}
\textbf{P}: \max_{\mathbf s}\quad C_{\scriptscriptstyle s} = R_{\scriptscriptstyle B} - R_{\scriptscriptstyle E}
\label{eq:optimization_problem}
\end{equation}
\begin{equation}
\text{s.t.} \quad 
s_{\scriptscriptstyle n}\ \in \{0,1\}, \qquad  \forall n \in N_{\scriptscriptstyle \text{irs}},
\label{eq:constraint1}
\end{equation}
\begin{equation}
\phantom{\text{s.t.} \quad 
s_{n}\in\{0,1\},} E(f) = 1, \qquad \forall f \in \left[0, \frac{1}{2T_{\scriptscriptstyle s}}\right].
\label{eq:constraint3}
\end{equation}
\end{subequations}
Where $C_{s}$ represents the achievable secrecy capacity of the wiretap channel, the constraint (\ref{eq:constraint1}) ensures that a single \ac{irs} element cannot be allocated to more than one user at a time, and constraint (\ref{eq:constraint3}) ensures a uniform power distribution over the frequency range. The optimisation problem, as defined in equation (\ref{eq:optimization_problem}), is inherently non-linear and non-convex due to the binary nature of allocation. This binary nature results in a combinatorial explosion of the search space, comprising $2^{N_{\scriptscriptstyle{\text{irs}}}}$ potential solutions. 

In order to demonstrate the effectiveness of the proposed \ac{pls} method in a time-efficient manner, we employ a \ac{ga}-based optimisation method to navigate this expansive search space. The \ac{ga} iteratively searches for the best user-element allocation, leveraging evolutionary processes to efficiently converge to high-quality solutions. The \ac{ga} is a powerful heuristic technique inspired by fitness improvement through evolution in biological systems, which has shown efficiency in solving optical wireless communication problems \cite{Ref9,GA:1}.  It commences with the generation of a diverse solution set, which represents the initial population in the optimisation landscape. Subsequently, it enhances these solutions through a systematic cycle of genetic operations, namely:  selection, which privileges the fittest solutions; crossover, which combines attributes of parent solutions; and mutation, which introduces necessary diversity, thereby propelling the population towards increasingly optimised solutions. The \ac{ga} ability to handle binary decision variables makes it particularly well-suited for problems like ours. By optimising the IRS elements' allocation, we can significantly reduce the complexity of the problem compared to \ac{es}. This approach enables the discovery of near-optimal solutions within a practical timeframe, making it a valuable tool for enhancing \ac{pls} in \ac{irs}-assisted VLC systems. The detail steps for the \ac{ga} are shown in Algorithm\ref{alg:GA}.

\begin{figure}[!t]
\centering
\includegraphics[width=3.5in]{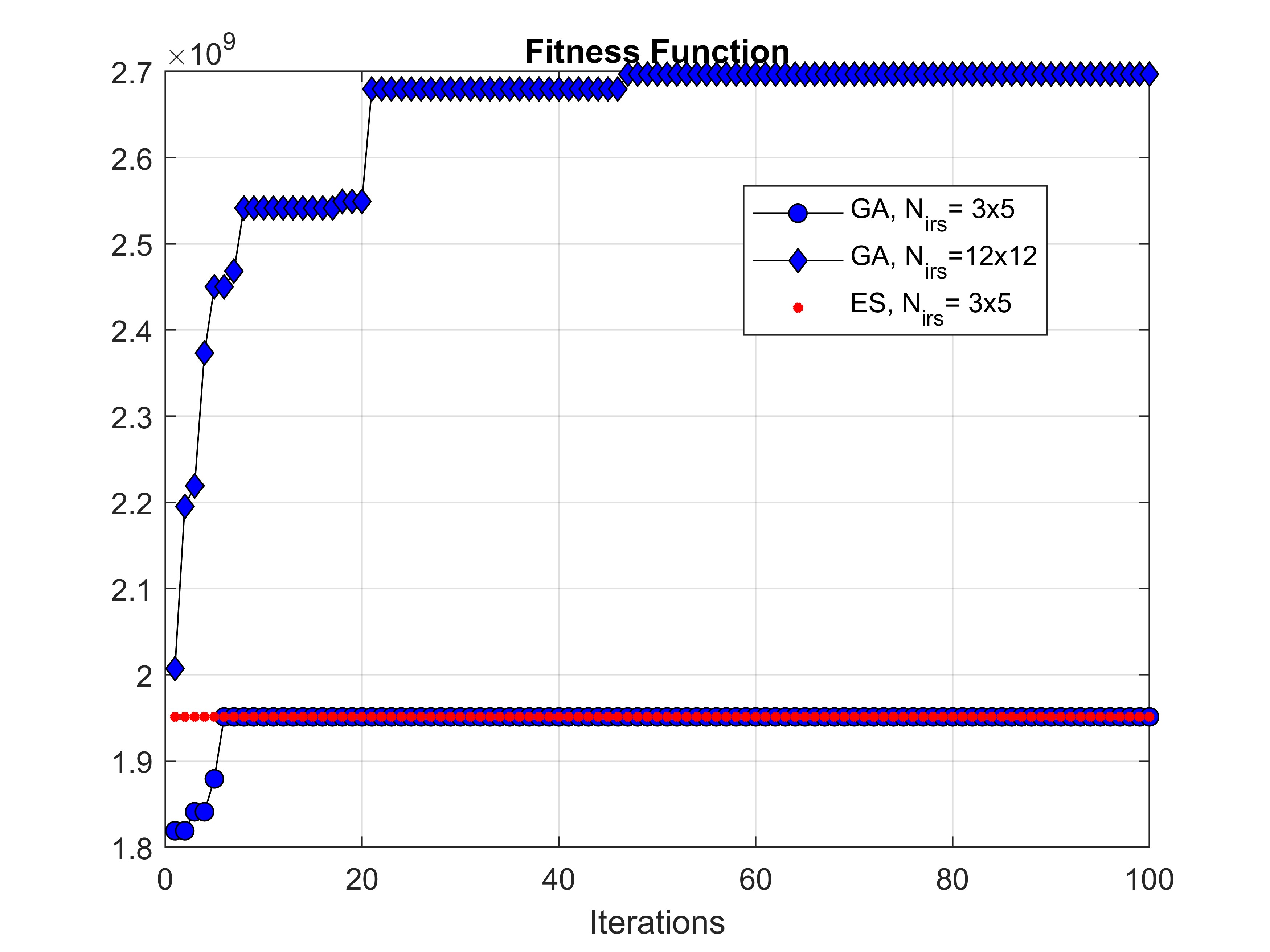}
\caption{Convergence rate of Algorithm \ref{alg:GA} for different IRS sizes.}
\label{fig_Ga}
\end{figure}

The computational complexity in Algorithm \ref{alg:GA}, which maximise the objective function (\ref{eq:optimization_problem}) subject to (\ref{eq:constraint1}) and (\ref{eq:constraint3}), is primarily determined by the population size \(S_p\) and the number of generations \(\mathbb{N}_{\scriptscriptstyle\text{Gen}}\). Each evaluation of the fitness function involves assessing the binary vector \(s\) of length \(N_{\scriptscriptstyle \text{irs}}\), where  \(N_{\scriptscriptstyle \text{irs}}\) is the total number of elements being allocated. Therefore, the complexity of evaluating the entire population in one generation is \(O(S_p \times N_{\scriptscriptstyle \text{irs}})\), and the total complexity over all generations becomes \(O(\mathbb{N}_{\scriptscriptstyle\text{Gen}}\times S_p \times N_{\scriptscriptstyle \text{irs}})\). Fig. \ref{fig_Ga} shows the convergence of  Algorithm \ref{alg:GA}. It is clear that the algorithm approaches the best solution within 6 iterations for $N_{\scriptscriptstyle\textbf{irs}}= 3\times5$, and within 47 iterations for $N_{\scriptscriptstyle\textbf{irs}}= 12\times12$. Utilising \ac{es} for the case of $N_{\scriptscriptstyle \text{irs}}$= $3 \times 5$ is possible, increasing to $N_{\scriptscriptstyle\textbf{irs}}= 12\times12$ makes \ac{es} very computationally expensive. This rapid convergence highlights the algorithm’s capability to effectively navigate the solution space and achieve near-optimal results swiftly. Since we achieve optimal performance compared to \ac{es} for a small number, it is an indication that the \ac{ga} is reliable in achieving an optimal or near-optimal solution.
\begin{algorithm}
\caption{GA-Based Maximisation of Secrecy Capacity}
\label{alg:GA}
\begin{algorithmic}[1] 
\Statex \textbf{Input:} $S_{\scriptscriptstyle p},\mathbb{N}_{\scriptscriptstyle\text{Gen}}$, channel gain:
\Statex \(g_{\scriptscriptstyle {B,\text{LED}}}^{\scriptscriptstyle \text{LoS}}\)
\Statex \(g_{\scriptscriptstyle{ E, \text{LED}}}^{\scriptscriptstyle \text{LoS}}\)
\Statex \(\textbf{g}_{\scriptscriptstyle{B,\text{LED}}}^{\scriptscriptstyle \text{NLoS}}\)=\([g_{\scriptscriptstyle {B,1,\text{LED}}}^{\scriptscriptstyle \text{NLoS}}, g_{\scriptscriptstyle {B,2,\text{LED}}}^{\scriptscriptstyle \text{NLoS}}, \dots, g_{\scriptscriptstyle {B,N_{\text{irs}},\text{LED}}}^{\scriptscriptstyle \text{NLoS}} ]\)
\Statex \(\textbf{g}_{\scriptscriptstyle{E,\text{LED}}}^{\scriptscriptstyle \text{NLoS}}\)= \([g_{\scriptscriptstyle{E,1,\text{LED}}}^{\scriptscriptstyle \text{NLoS}}, g_{\scriptscriptstyle{E,2,\text{LED}}}^{\scriptscriptstyle \text{NLoS}},\dots, g_{\scriptscriptstyle{E,N_{\text{irs}},\text{LED}}}^{\scriptscriptstyle \text{NLoS}}]\)
\Statex \textbf{Output:} Binary allocation vector $s$ maximising (\ref{eq:optimization_problem})
\State Generate a random initial population of size $S_p$
\State Evaluate the initial population using (\ref{eq:optimization_problem})
\State Set generation count $g = 1$
\While{$g \leq \mathbb{N}_{\scriptscriptstyle\text{Gen}}$}
    \State Select parents from the population based on fitness
    \State Apply crossover with probability $P_c$ to create offspring
    \State Apply mutation with probability $P_m$ to offspring
    \State Evaluate new candidates using (12a)
    \State Select the best individuals for the next generation
    \State $g = g + 1$
\EndWhile
\State \textbf{return} Binary allocation vector $s$
\end{algorithmic}
\end{algorithm}
\setlength{\abovecaptionskip}{5pt}

\section{Simulation Results} 
\setlength{\abovecaptionskip}{5pt}
{\small \begin{table}[t!]
\caption{Simulation Parameters}
\label{table:simulation_parameters}
\centering
\begin{tabular}{|c|c|}
\hline
\textbf{Parameters} & \textbf{Values} \\
\hline
LED half-power semi-angle & \( \Phi_{\scriptscriptstyle 1/2} = 60^\circ \) \\
\hline
PD responsivity [A/W] & \( R_{\scriptscriptstyle \text{pd}} = 0.6 \) \\
\hline
PD physical area [cm\(^2\)] & \( A_{\scriptscriptstyle d} = 1 \) \\
\hline
PD FoV & \( \Psi_{\scriptscriptstyle \text{FoV}} = 90^\circ \) \\
\hline
Symbol period [ns] & \( T_s = 1 \) \\
\hline
Noise power spectral density &  \(\mathcal{N} = 10^{-21}\) \\
\hline
IRS reflectivity & \( \rho = 1 \) \\
\hline
Refractive index of PD & \(\xi = 1.5 \) \\
\hline
Modulation scaling factor & \( \mathsf{k} = 3.2 \) \\
\hline
Gap value [dB] & \( \Gamma = 2 \) \\
\hline
IRS size & \( N_{\scriptscriptstyle \text{irs}}=N_x \times N_y = 12 \times 12 \) \\
\hline
Separation distance [cm] & \( D=30  \) \\
\hline
Population size & \( S_{\scriptscriptstyle p}=50  \) \\
\hline
Number of generations &\(\mathbb{N}_{\scriptscriptstyle\text{Gen}}=30\)\\
\hline
Crossover Probability          & \( P_c = 0.8 \) \\
\hline
Mutation Probability           & \( P_m = \frac{1}{N_{\text{irs}}} \) \\ \hline
\end{tabular}
\label{table 1}
\end{table}}
\begin{figure}[!t]
\centering
\includegraphics[width=3.5in]{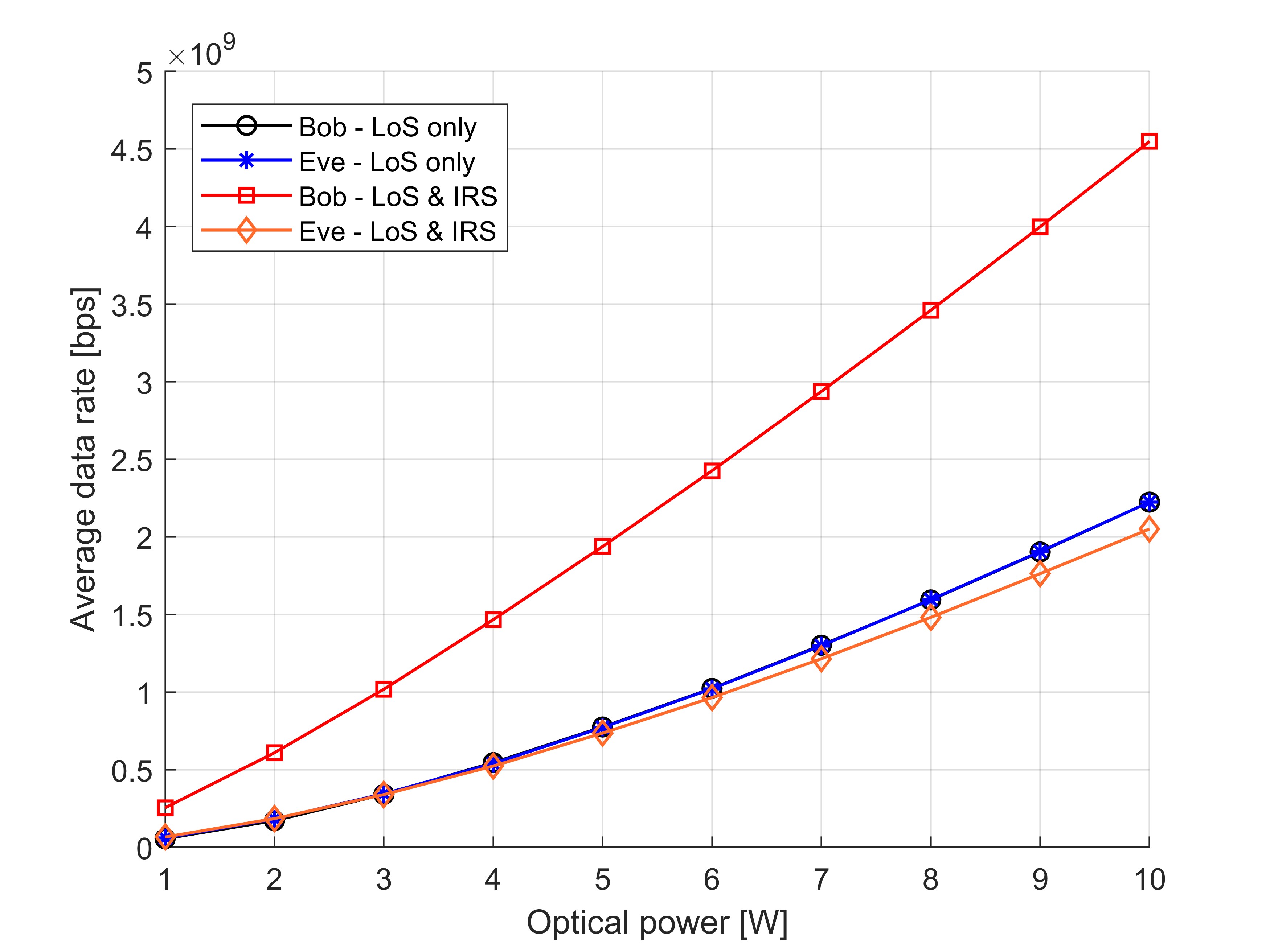}
\caption{The average data rate of Bob and Eve for random locations across the room.}
\label{fig_sub1}
\end{figure}
\begin{figure}[!t]
\centering
\includegraphics[width=3.5in]{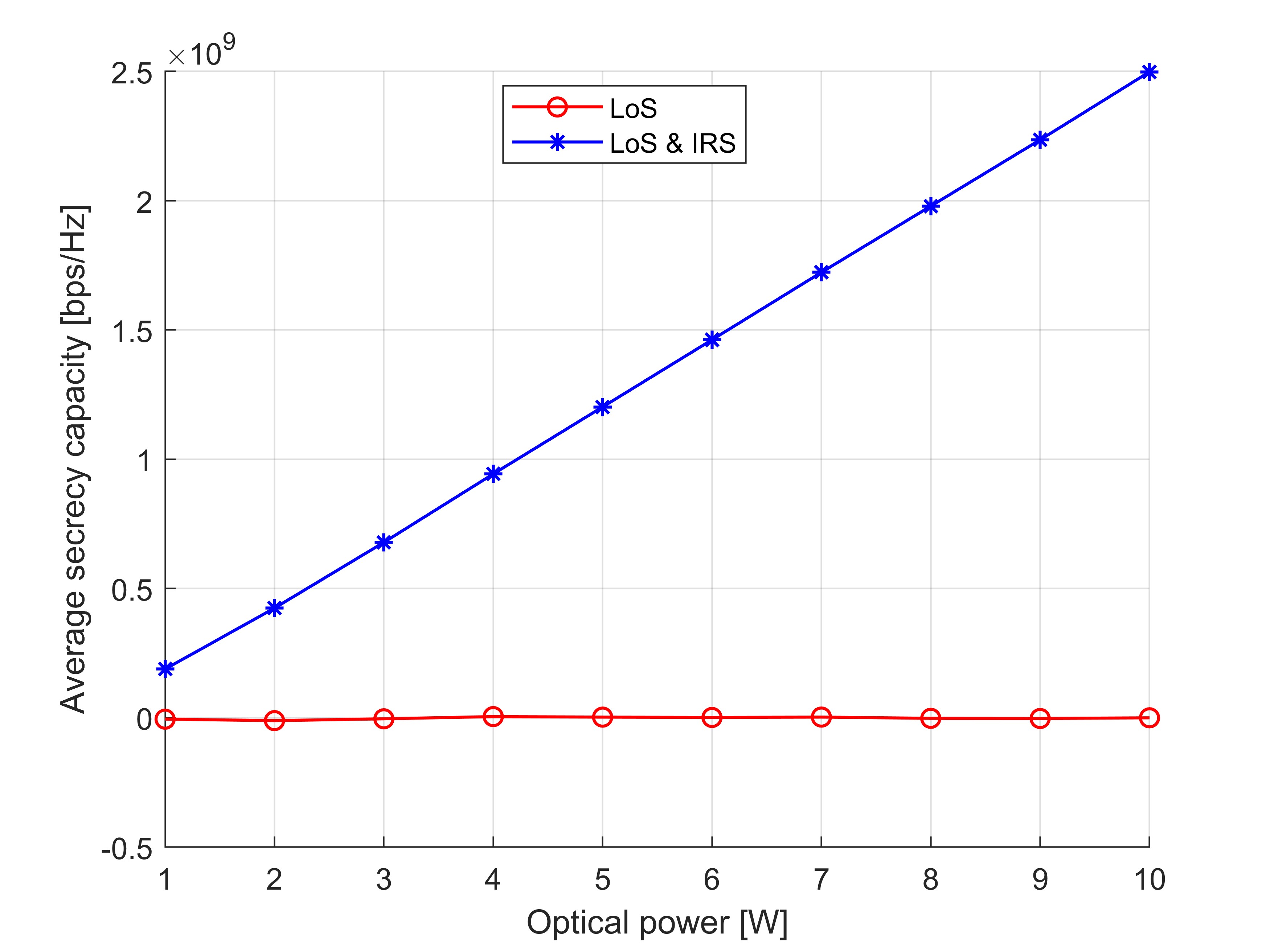}
\caption{ Average secrecy capacity for uniformly distributed location.}
\label{fig_sub2}
\end{figure}
\begin{figure}[!t]
\centering
\includegraphics[width=3.5in]{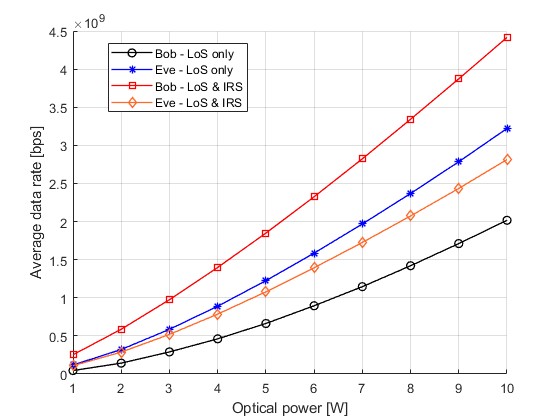}
\caption{Average data rate at random positions with Eve positioned within the LED's inner zone}
\label{fig_sub3}
\end{figure}
\begin{figure}[!t]
\centering
\includegraphics[width=3.5in]{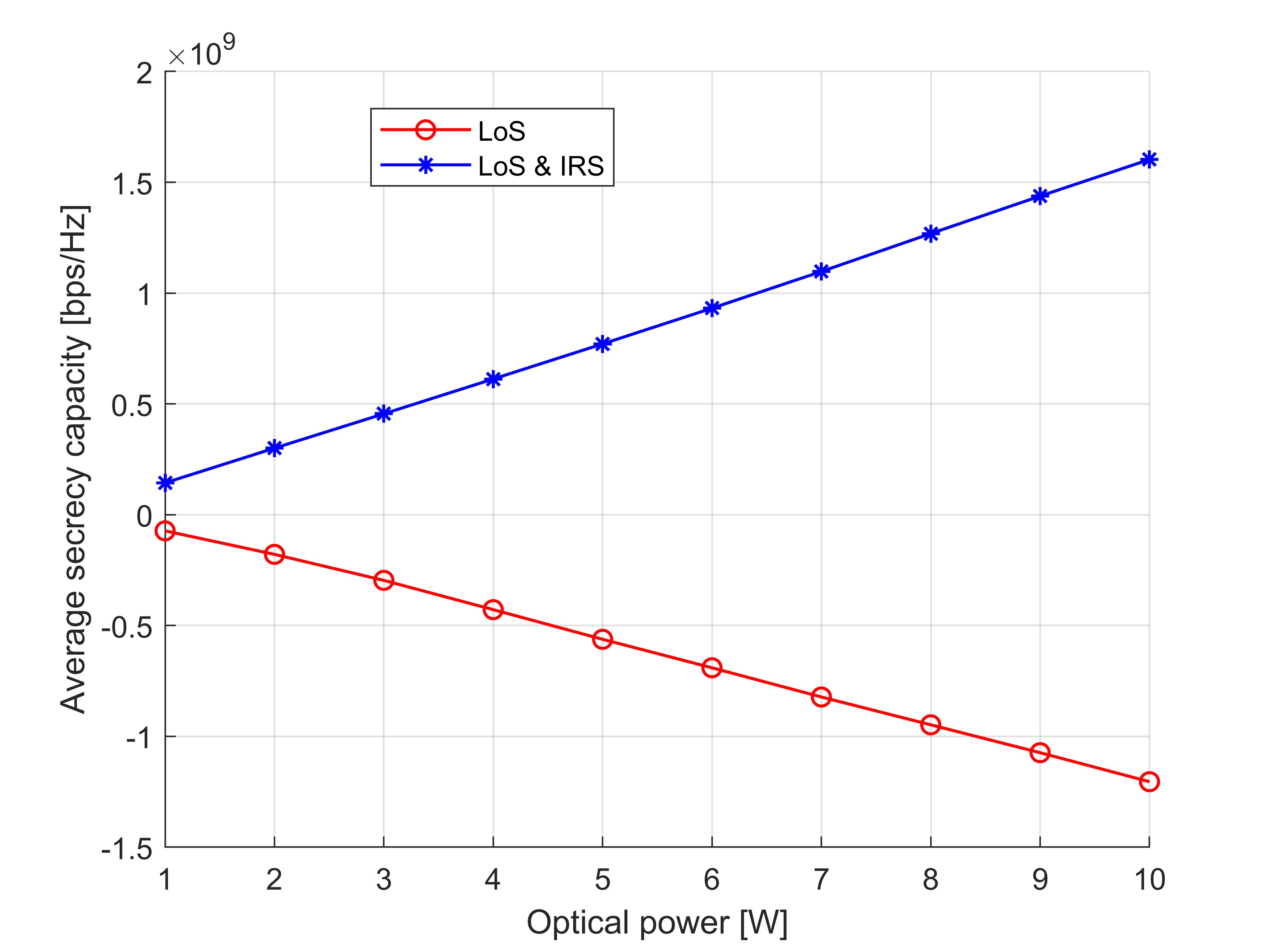}
\caption{Secrecy capacity at random positions with Eve positioned within the LED's inner zone.}
\label{fig_sub4}
\end{figure}

In this section, we discuss the simulation results of an indoor room with dimensions of $5m\times5m\times3m$. The simulation parameters are shown in Table \ref{table 1}. The main goal of these simulations is to demonstrate the effect of time delay, which leads to constructive and destructive \ac{isi}, and the effectiveness of the proposed \ac{pls} mechanism.

In Fig.~\ref{fig_sub1}, a comparison of the data rate against different power levels for Bob and Eve users with and without \ac{irs} is shown. The users are located randomly, and the simulation runs for 300 iterations, with averages taken. The results demonstrate how the use of \ac{irs} can improve the performance of Bob by allocating elements that create constructive \ac{isi}. Conversely, it is also shown that elements that cause the worst delay and create destructive \ac{isi} can degrade the signal power for Eve. Fig. \ref{fig_sub2} shows the corresponding secrecy capacity performance. It is noted that the secrecy capacity remains at zero for the \ac{los} case since both users are randomly located following uniform distribution, so the average difference in their rate values remains zero. Moreover, it is noted that the enhancement in Bob's data rate, evident from Fig. \ref{fig_sub1}, can result in a significant increase in the secrecy capacity even though Eve's rate is not highly affected. 

In the following, we investigate the performance when the conditions of Bob's and Eve's locations are not equal. This is for a better demonstration of the effect of the destructive \ac{isi} in enhancing the secrecy capacity in certain scenarios. We divide the coverage area into two zones: an inner zone with a radius of \(1 \, \text{m}\) within the \ac{led} and an outer zone that includes areas further than \(1 \, \text{m}\) from the \ac{led} in radius. We assume a worst-case scenario in which Eve is closer to the \ac{led} than Bob. The simulation results in Fig.~\ref{fig_sub3} show how the time delay and the associated \ac{isi} can enhance Bob's performance and degrade Eve's performance. It is noted that at \(7 \, \text{W}\) optical power, the rate at Bob improved by \(59.41\%\), while the rate at Eve dropped by \(12.30\%\). Fig.~\ref{fig_sub4} shows the corresponding secrecy capacity for the worst-case scenario when Eve is located in the inner zone and Bob in the outer zone. It is noted that secrecy capacity is negative for the \ac{los} case because Eve has more favourable channel conditions that Bob due to its proximity to the LED. We can see a significant enhancement of \(253.7 \%\) at \(3 \, \text{W}\) power, where the average secrecy capacity of \ac{los} is \( - 296.171 \, \text{Mbps/Hz} \), and the average secrecy capacity of the combined \ac{los} and \ac{irs} is \( 45.5398 \, \text{Mbps/Hz} \). Such enhancement is usually not possible with conventional \ac{pls} used in literature when Eve is positioned closer to the \ac{led}.
\balance
\section{Conclusion}
This paper presents a novel approach that utilises the time delay aspect in \ac{irs}-assisted VLC systems for establishing \ac{pls}. Our results show that a simple \ac{ga}-based \ac{irs} allocation can result in significant secrecy capacity enhancement, even when the eavesdropper is closer to the \ac{led} than the legitimate user. We believe that further investigations of \ac{irs}-induced time delay and the utilisation of constructive and destructive \ac{isi} patterns can open the door for equipping VLC with an even higher security advantage against interception attacks.  
\small
\balance

\end{document}